\documentclass[a4paper,11pt]{article}
\pdfoutput=1 

\usepackage{jheppub} 

\usepackage[T1]{fontenc} 

\usepackage{graphicx}  
\usepackage{dcolumn}   
\usepackage{bm}        
\usepackage{amssymb, nccmath}   
\usepackage{graphicx, epsfig, subfigure}
\graphicspath{{tmpfig/}{figures/}}

\usepackage{amsmath}
\usepackage{slashed}
\usepackage{lipsum}
\usepackage{subfigure}
\usepackage{mathrsfs}
\usepackage{longtable}
\usepackage{rotating}
\usepackage{mathtools}
\usepackage{cuted}
\usepackage{color}
\usepackage{lineno}

\usepackage{setspace}
\title{Search for Axion(-like) Particles in Heavy-Ion Collisions}

\author[]{Yi Yang, \note{Corresponding author.}} 
\author[]{and Cheng-Wei Lin} 

\affiliation[]{Department of Physics, National Cheng Kung University, Tainan, 70101, Taiwan, ROC}

\emailAdd{yiyang@ncku.edu.tw}

\abstract{    We propose a novel way to search for axion(-like) particles in heavy-ion collisions using prompt photons as the probe and the property of conversion between photon and axion(-like) particles under a strong magnetic field generated in the non-central collisions. 
    The expected result reveals that a new phase space region of the coupling constant for photon and axion(-like) particles can be covered in the future high energy nuclear colliders. 
    }

\begin{document} 
\maketitle
\flushbottom

\section{Introduction}
The Standard Model of Particle Physics (SM) is the mathematical framework to describe the interactions between elementary particles including electromagnetism, weak, and strong interactions. 
In the past two decades, we had tremendous success on understanding of the SM from many important experiments, for instance the experiments at the Relativistic Heavy Ion Collider (RHIC) at Brookhaven National Laboratory, STAR and PHENIX, which devoted themselves into understanding the new state of matter of Quantum Chromodynamics (QCD), Quark-Gluon Plasma (QGP), since 2000~\cite{rhic_qgp_1, rhic_qgp_2}, the experiments at the Large Hadron Collider at CERN, ATLAS and CMS, which discovered the Higgs boson in 2012~\cite{lhc_higgs_1, lhc_higgs_2} and provided many precision measurements on the electroweak sector~\cite{cern_ew}, and many other experiments not mentioned here. 
However, there are still many unsolved puzzles in physics, such as the origin of matter-antimatter asymmetry~\cite{Sakharov}, the dark matter~\cite{dark_matt}, and dark energy~\cite{dark_eng} problem.

One of the most intriguing mysteries is that the missing mechanism of the Charge-Parity ($CP$) violation processes in the strong interactions, known as the ``strong $CP$ problem in QCD''. 
The missing $CP$ violation processes are one of missing ingredients in the ``Sakharov conditions''~\cite{Sakharov} to explain the matter-antimatter asymmetry problem in our universe.  
To overcome the strong $CP$ problem, a new mechanism was proposed by Roberto Peccei and Helen Quinn in 1977 by adding an extra global U(1) gauge symmetry in the Lagrangian~\cite{peccie_quinn}.
One year later, Steven Weinberg~\cite{weinberg} and Frank Wilczek~\cite{wilczek} implemented the breaking of this new U(1) gauge symmetry and predicted a new hypothetical spin-0 pseudoscalar particle, axion. 

Recently, many theories also predict very light pseudoscalar or scalar particles, which have very similar properties but play no parts in solving the strong $CP$ problem, so called axion-like particles. 
Both axion or axion-like particles must be weakly interacting with normal particles. 
Therefore, these particles are the perfect candidates to solve the dark matter problem~\cite{wisp, dm_1, dm_2, dm_3}.
There are many experimental constrains on the coupling strength and the mass of axion(-like) particles from low energy nuclear physics, high energy particle physics, and astrophysics. 
Some experiments are based on the property of conversion between photon and axion(-like) particles to probe extremely low mass regions and they will be discussed again later. 
Some approaches use collider signatures to search for axion(-like) particles, and they can cover the mass region from 0.1 to 100 GeV, for example using photon-jet as a probe in $p$+$p$ collisions~\cite{photon_jet}, or relying on the strong electromagnetic field generated by the ultra-peripheral heavy-ion collisions~\cite{upc_1, upc_2} and this has already been tested in the LHC~\cite{cms}.

In this paper, we propose an alternative way to search for the axion(-like) particles signature using the prompt photon production in heavy-ion collisions.  

\section{Photon-Axion(-like) Conversion}
\begin{figure}[!htbp]
  \begin{center}
      \includegraphics[width=0.45\textwidth]{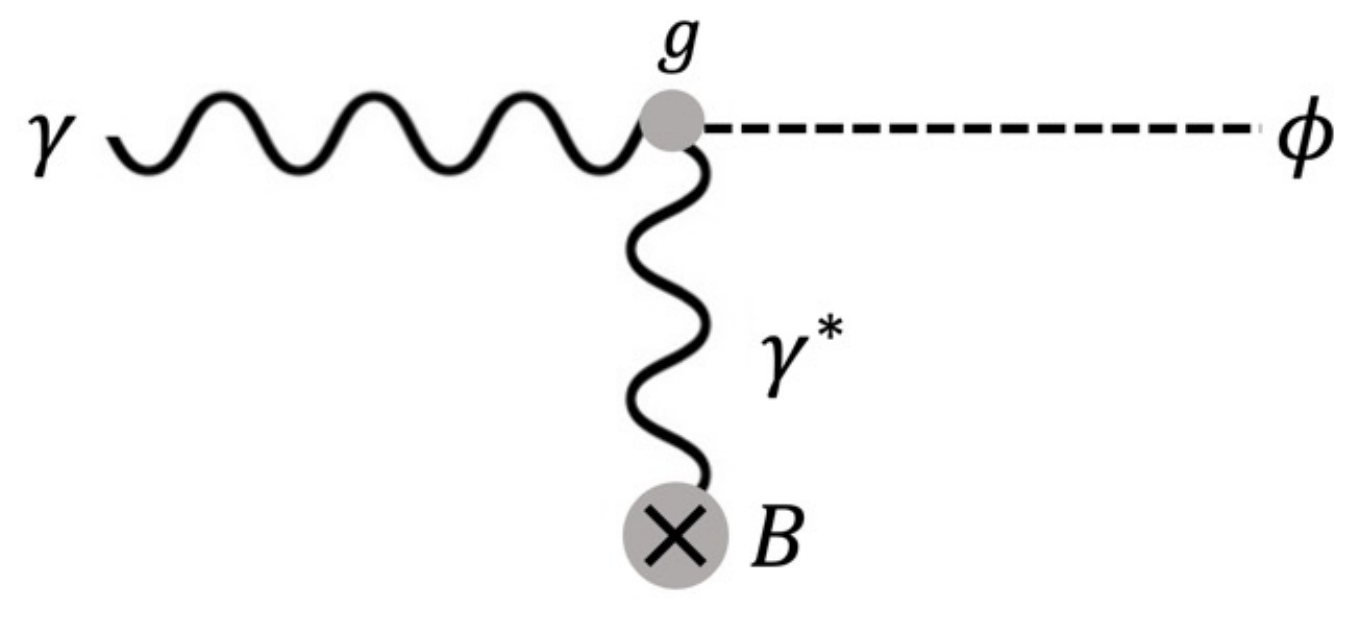}
      \end{center}
    \caption{ The Feynman-like diagram of photons coupling to axion(-like) particles via a weak coupling constant $g$.  
  \label{fig:fey_ph_alps}}
\end{figure}

One of the most interesting features of axion(-like) particles is that they can couple to photon via a weak-strength coupling constant, $g$.
Figure~\ref{fig:fey_ph_alps} shows the Feynman-like diagram of the conversion of photons to axion(-like) particles via the interaction with the magnetic field ($B$) and this corresponds to the non-resonant $\gamma + \gamma^* \to \phi$ production, where $\gamma^*$ is the photon from the $B$ field.

The probability of photon to axion(-like) particles or axion(-like) particles to photon conversion can be derived directly and many review articles have detailed description of it (see Ref.~\cite{review_1, review_2}).
The conversion probability can be written as:
\begin{eqnarray}
    \label{eq:prob_gamma_to_phi}
    P^{\gamma \to \phi} &=& 4\frac{g^2 B^2 \omega^2}{m_{\phi}^4}\sin^2\left(\frac{m_{\phi}^2 L}{4\omega}\right)   \\  
    &\approx& \left(\frac{gBL}{2}\right)^2 \ \ \ \ {\rm if \ } m_{\phi}^2 L/4\omega \ll 1, \nonumber
\end{eqnarray}
where $L$ is the interaction distance of the photon or axion(-like) particles with the magnetic field $B$ field, $\omega$ is the photon's frequency, and $m_{\phi}$ is the rest mass of axion(-like) particles.
An interesting idea of searching for axion(-like) particles using this photon-axion(-like) conversion, so called light-shining-through-walls (LSW), was proposed~\cite{lsw_1, lsw_2, lsw_3}. 
The basic idea is to use the possibility of photon converting to axion(-like) particles under a strong magnetic field and then axion(-like) particle will penetrate through the wall due to the extremely weak interaction between axion(-like) particles and the SM particles (the Wall). 
Finally, the axion(-like) particles will convert back to photon under a strong magnetic field and then to be detected. 
The total probability of the LSW experiments is the product of two conversion probabilities as followings: 
\begin{eqnarray}
    \label{eq:prob_gamma_to_phi_to_gamma}    
    P^{\gamma \to \phi \to \gamma} &=& P^{\gamma \to \phi} \times P^{\phi \to \gamma} \nonumber \\ 
    &\approx& \left(\frac{gBL}{2}\right)^4.    
\end{eqnarray}
Many experiments are dedicated on the LSW-type search, such as ALPS~\cite{alps_1, alps_2} and BFRT~\cite{bfrt}. 
The results mainly focused on the phase space in the very low-mass region, namely $< 1$ eV, and the upper limit on $g$ is around $10^{-7}$ to $10^{-6}$ GeV$^{-1}$. 
An important observation from Eqs.~\ref{eq:prob_gamma_to_phi} and~\ref{eq:prob_gamma_to_phi_to_gamma} is that the conversion probability also depends on the product of $B$ and $L$, the $BL$ factor. 
In other words, the experimental sensitivity for LSW on the coupling constant $g$ is inverse proportional to $BL$ and proportional to $(P^{\gamma \to \phi \to \gamma})^{-1/4}$. 
Therefore, the current results from the LSW-type experiments are limited by the $BL$ factor which is in the order of 10 to 100 $T\cdot m$\footnote{The ALPS experiment has 5 $T$ magnetic field for 8.8 $m$ ($BL$ = 44 $T\cdot m$) and the OSQAR experiment has 9 $T$ magnetic field for 7 $m$ ($BL$ = 63 $T\cdot m$).}~\cite{alps_1,osqar}.

\section{Search for axion(-like) particles in heavy-ion collisions} 
\begin{figure}[!htbp]
  \begin{center}
      \includegraphics[width=0.35\textwidth]{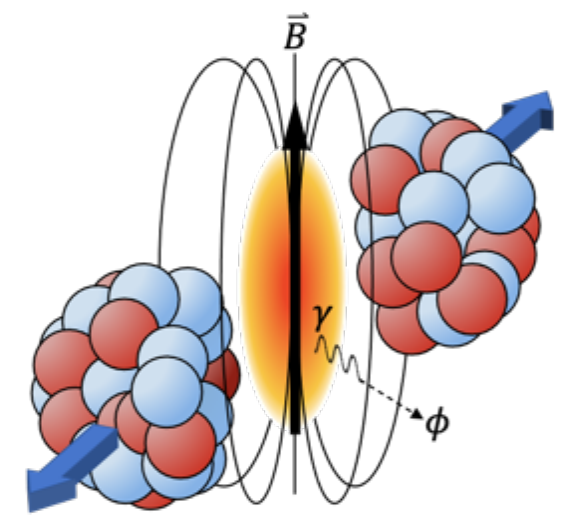}
      \end{center}
    \caption{ The illustration of the prompt photon produced in heavy-ion collisions and it converts to axion(-like) particle inside the fire ball which has strong magnetic field.  
  \label{fig:photon_axion_in_fireball}}
\end{figure}

As mentioned previously, two key components for observing the conversion between photon and axion(-like) particles are the $BL$ factor and the photon source. 
In the relativistic heavy-ion collisions, a very strong magnetic field can be generated in the non-central collisions due to the large charges and high speed of the colliding nuclei. 
The strength of the magnetic field in the interaction area can be estimated as $eB \sim f m_{\pi}^2$, where $m_{\pi}$ is the mass of $\pi$ meson and $f$ is the scaling factor which depends on the type and energy of collisions particles~\cite{mag_field, larg_mag_field}.
At the top of the RHIC collision energy, $\sqrt{s_{\rm NN}} = 200$ GeV in Au+Au collisions, the magnetic field can be as high as $4 \times 10^{14}\ T$ with the impact parameter $b = 10$ fm and it is much stronger than any apparatus in labs~\cite{mag_time}. 
More importantly, the strength of magnetic field is linearly proportional to the collision energy~\cite{mag_vs_eng}, as
\begin{eqnarray}
    \label{eq:mag_vs_energy}
     B(\sqrt{s_{\rm NN}}) = \frac{\sqrt{s_{\rm NN}}}{200\ {\rm GeV}} \times 4 \times 10^{14}\ T.
\end{eqnarray}
However, this extremely strong magnetic field will be disappeared rapidly, in the time scale of 1 fm/$c$, where $c$ is the speed of light~\cite{mag_time}. 
Fortunately, this strong magnetic field can be very uniform in the time scale of 0.1 fm/$c$ which corresponding to $L = 0.1\ {\rm fm}/c \times c = 10^{-16}\ m$ and therefore Eq.~\ref{eq:prob_gamma_to_phi} is valid in this short period of time. 
Since the precise time-evolution is very complicated to estimate if the medium effect from QGP is considered~\cite{mag_vs_eng}, a rough time scale of 0.1 fm/$c$ is used for the future discussion. 
Consequently, the $BL$ factor in heavy-ion collisions can also reach up to 20$-$50 $T\cdot m$ at future collider energies and it will be described later. 
On the other hand, the photons produced in heavy-ion collisions can be good candidates for the search of axion(-like) particles by taking the advantage that photons have the chance to convert to axion(-like) particles in the aforementioned strong magnetic field, as illustrated in Fig.~\ref{fig:photon_axion_in_fireball}. 
Note that there are three kinds of photons generated in heavy-ion collisions: prompt photons, thermal photons, and decay photons. 
The first two are normally categorized as the ``direct photons''. 

The probability of photon converting to axion(-like) particles in heavy-ion collisions can be extracted by comparing the prompt photon productions in $A$+$A$ collisions to that in $p$+$p$ collisions, and coincidentally this is the nuclear modification factor $R_{\rm AA}^{\gamma}$ of the prompt photon. 
The $R_{\rm AA}^{\gamma}$ is defined as 
\begin{eqnarray}
    R_{\rm AA}^{\gamma} = \frac{1}{\langle N_{coll} \rangle}\frac{\left(\frac{dN^{\gamma}}{dX}\right)_{A+A}}{\left(\frac{dN^{\gamma}}{dX}\right)_{p+p}},    
\end{eqnarray}
where $\langle N_{coll} \rangle$ is the average number of binary nucleon-nucleon collisions in a given centrality bin, $\left(\frac{dN^{\gamma}}{dX}\right)_{A+A}$ and $\left(\frac{dN^{\gamma}}{dX}\right)_{p+p}$ are differential invariant yields of prompt photon in $A$+$A$ and $p$+$p$ collisions with respect to a certain kinematic variable $X$, respectively. 

The prompt photon production is expected to be not affected by the QGP medium since photon doesn't carry any color charge and there are two additional effects should be taken into account in the $R_{\rm AA}^{\gamma}$ determination. 
The first one is the conversion probability of photon to axion(-like) particles, and the second one is the axion(-like) particles produced in $A$+$A$ collisions then converting to photon. 
However, the production cross section for axion(-like) particles is also known to be extremely small, otherwise they have already been discovered in other experiments, so the second term can be ignored. 
The modified $R_{\rm AA}^{\gamma}$ should be rewritten as
\begin{eqnarray}
    R_{\rm AA}^{\gamma} &=&  \frac{ \left(\frac{dN^{\gamma}}{dX} \times (1 - P^{\gamma \to \phi}) + \frac{dN^{\phi}}{dX} \times P^{\phi \to \gamma} \right)_{A+A} }{ \langle N_{coll} \rangle \times \left( \frac{dN^{\gamma}}{dX}\right)_{p+p} }  \nonumber \\
    &\approx& \left( \frac{1}{\langle N_{coll} \rangle} \frac{ \left(\frac{dN^{\gamma}}{dX}\right)_{A+A} }{ \left(\frac{dN^{\gamma}}{dX}\right)_{p+p} } \right) \times (1 - P^{\gamma \to \phi}), 
\end{eqnarray}
where $\frac{dN^{\phi}}{dX}$ is the production yields for axion(-like) particles in $A$+$A$ collisions and this can be ignored due to the small production cross section as mentioned before. 
On the other hand, the electron-positron pair production from a single photon in a strong magnetic field might also contribute to the reduction of prompt photon yield~\cite{ep_prod}. 
This process is dependent on the angle of the photon to the magnetic field and it can be reduced by considering the prompt photon produced along the event plane of heavy-ion collisions. 
The precise accuracy of the prediction on this contribution, namely a few percent level, will help interpret the results.   

Finally, the conversion probability of photon to axion(-like) particles can be determined by the precision of the measured $R_{\rm AA}^{\gamma}$. 
In other words, the conversion probability of photon to axion(-like) particles equals to the probability of $R_{\rm AA}^{\gamma}$ away from unity, namely $R_{\rm AA}^{\gamma} < 1$.
It is worthwhile to mention that this approach of searching for axion(-like) particles in heavy-ion collisions only depends on $g^2$, unlike the traditional LSW experiments which depends on $g^4$, so it will provide us higher probability to observe them.

\section{The Expected Results}
\begin{figure}[!htbp]
  \begin{center}
      \label{raa}
      \includegraphics[width=0.55\textwidth]{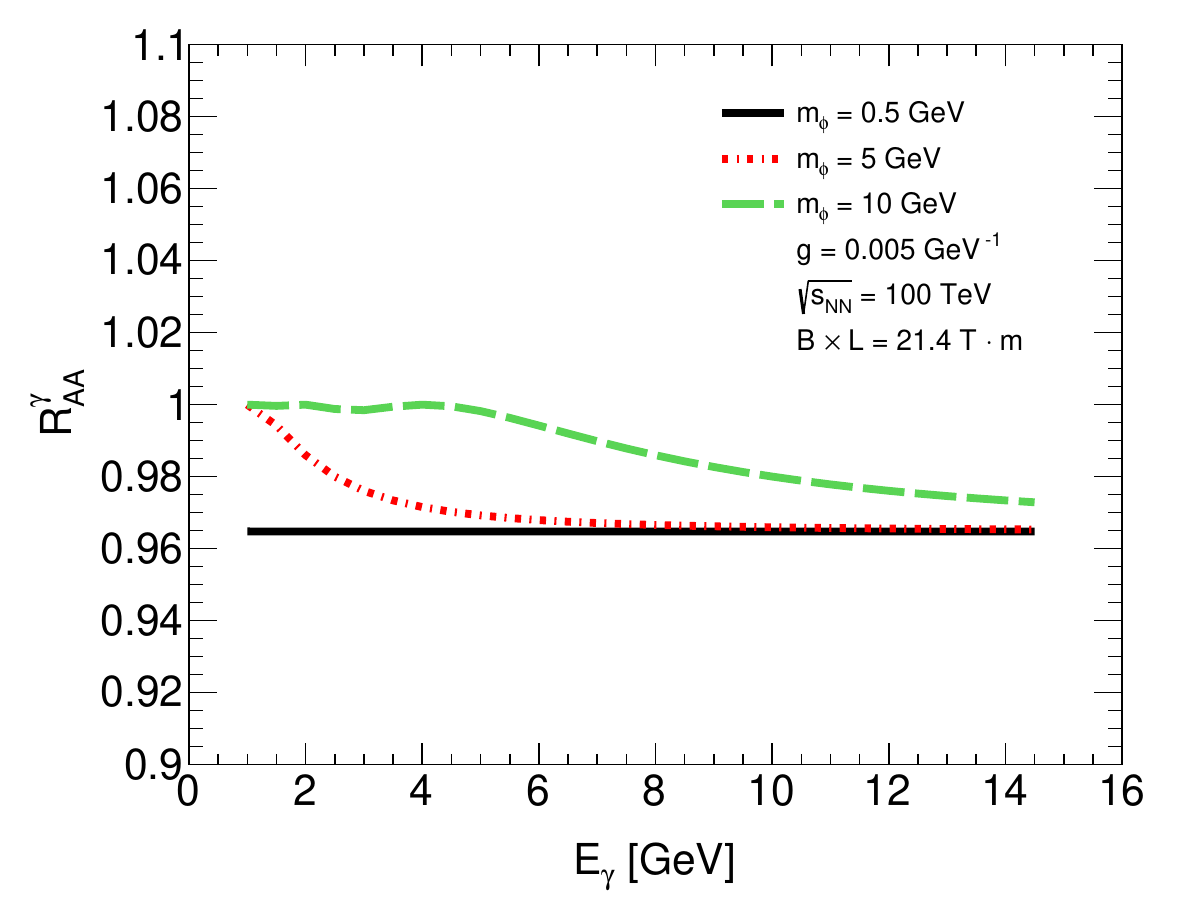}
      \end{center}
    \caption{ Example of the $R_{\rm AA}^{\gamma}$ of the prompt photon as a function of energy with $BL$ = 21.4 $T\cdot m$, $g = 0.005$ GeV$^{-1}$, and $m_{\phi} = 0.5$ GeV (black solid line), 5 GeV (red dotted line), and 10 GeV (green dashed line).  
  \label{fig:raa_results}}
\end{figure}

At the RHIC's top energy, $\sqrt{s_{\rm NN}}$ = 200 GeV in Au+Au collisions with the impact parameter $b = 10$ fm which corresponds to $30 - 60\%$ centrality~\cite{cent}, the $BL$ factor estimated from the previous section is only 0.04 $T\cdot m$ (even for the top LHC energy, $\sqrt{s_{\rm NN}}$ = 5.5 TeV in Pb+Pb collisions, the corresponding $BL$ factor is about 1 $T\cdot m$) and it is much smaller than the current LSW experiments. 
However, in the future high energy nuclear colliders, according to Eq.~\ref{eq:mag_vs_energy}, the $BL$ factor can reach 21.4 and 53.5 $T\cdot m$ for 100 TeV and 250 TeV Au+Au (or Pb+Pb, Xe+Xe) collisions, respectively. 
Also taking the advantage that this method only depends on $g^2$, instead of $g^4$ in the LSW-type experiments. 
Figure~\ref{fig:raa_results} shows the expected of the $R_{\rm AA}^{\gamma}$ of the prompt photon as a function of the photon energy with the configurations of $BL$ = 21.4 $T\cdot m$, $g = 0.005$ GeV$^{-1}$, and three different masses of the axion-(like) particles, $m_{\phi} = 0.5$, 5, and 10 GeV. 
It clearly shows that the $R_{\rm AA}^{\gamma}$ is deviated from unity if the $BL$ factor is large. 

The expected upper limit of the coupling constant of axion(-like) particle to photons, $g$, can be estimated by using the aforementioned $BL$ factors and the precision of the measured $R_{\rm AA}^{\gamma}$. 
However, in reality, there are also many other physics processes might affect the prompt photon yield in heavy-ion collisions, such as the contamination from the thermal photons, the nuclear parton distribution functions (nPDF) effect~\cite{npdf}, and the fragmentation photons from quarks or gluons traversing the plasma. 
Fortunately, these effects can be reduced significantly by some experimental treatments: (1) requiring the energy of direct photons to be larger than 5 GeV, based on the ALICE results~\cite{alice_dir_photon}, to avoid the thermal photons production and the nPDF effect which is known to be only significant at the low energy region; (2) focusing only on the isolated photons to reduce the contribution from the fragmentation photons. 
Furthermore, since analysis techniques are also expected to be improved in the foreseeable future, for example using machine (deep) learning architecture, the purity of the prompt photons candidates will also be improved significantly. 

\begin{figure*}[!htbp]
  \begin{center}
      \includegraphics[width=0.95\textwidth]{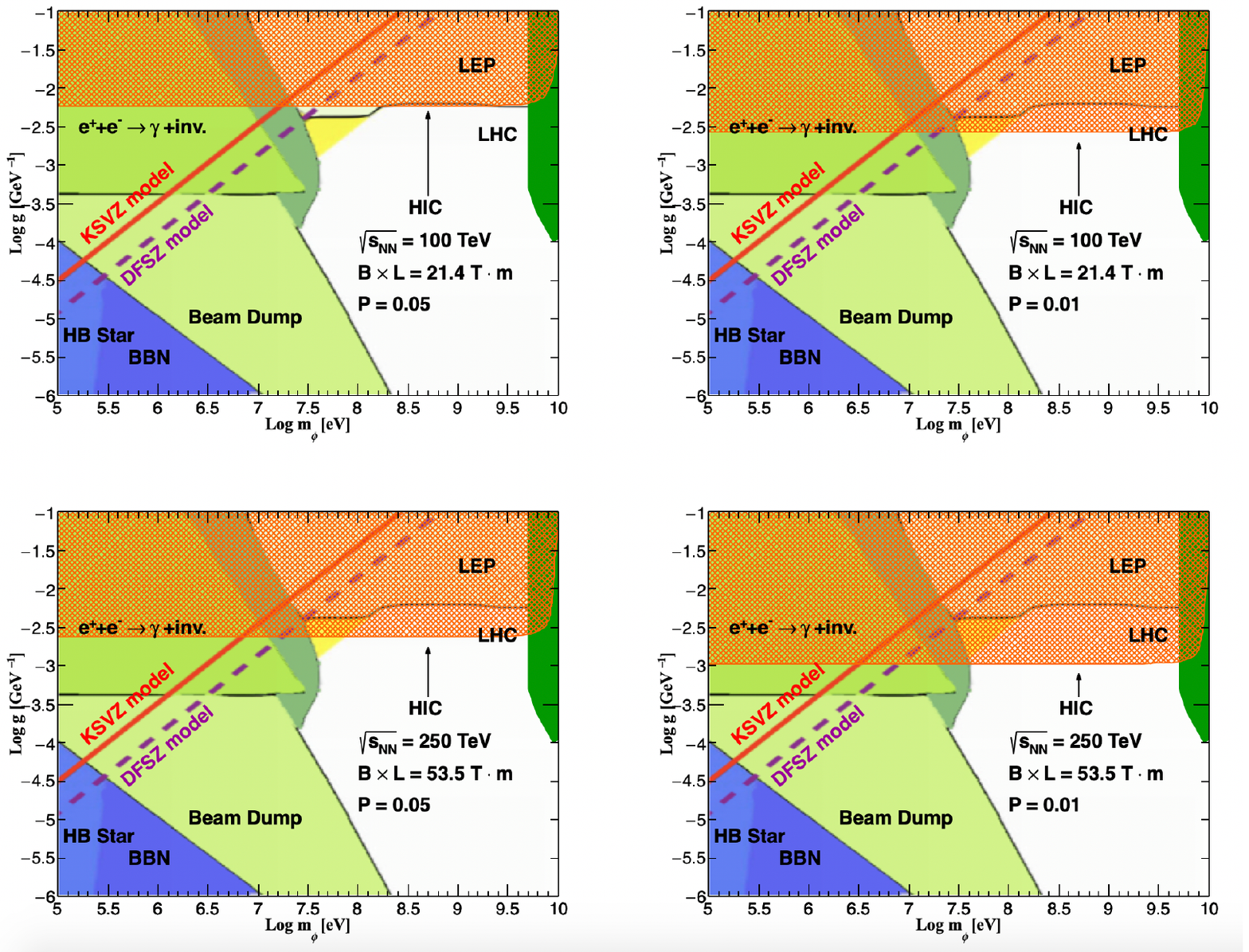}
      \end{center}
    \caption{ The expected upper limits on the coupling constant, $g$, as a function of the mass of axion(-like) particles with the different assumptions of the $BL$ factor and the probability ($P$) of the measured $R_{\rm AA}^{\gamma}$ away from unity are shown as orange-shaded area:
    (a) $BL = 21.4$ $T\cdot m$ and $P = 5\%$, (b) $BL = 21.4$ $T\cdot m$ and $P = 1\%$, (c) $BL = 53.5$ $T\cdot m$ and $P = 5\%$, and (d) $BL = 53.5$ $T\cdot m$ and $P = 1\%$. 
    The QCD axion models lie within an order of magnitude from the explicitly shown the ``KSVZ'' axion line~\cite{ksvz_1,ksvz_2} (red solid line) and ``DFSZ'' model (purple dashed line)~\cite{dfsz_1, dfsz_2}. The other colored regions are: experimentally excluded regions (green)~\cite{cms, ee_lep, bd_1, bd_2, bd_3,bd_4, na64_1, na64_2}, constraints from astrophysical or cosmological arguments (blue)~\cite{hb_1, hb_2, bbn_1, bbn_2}. 
  \label{fig:results}}
\end{figure*}

Figure~\ref{fig:results} shows the estimated upper limits of $g$ as a function of the mass of axion(-like) particles with the assumptions of $BL$ equals to 21.4 or 53.5 $T\cdot m$ and the probability of the measured $R_{\rm AA}^{\gamma}$ away from unity equals to $1\%$ or $5\%$ for demonstration.
The excluded region stops beyond $m_{\phi} \sim$ 10 GeV is due to the photon energy ($\omega$) in Eq.~\ref{eq:prob_gamma_to_phi} is set to be 5 GeV. 
Note that one can also consider a different kind of nuclear modification factor $R_{\rm CP}$ to improve the precision. 
The $R_{\rm CP}$ is defined as the ratio of the invariant yield in the head-to-head (central) heavy-ion collisions and the invariant yield in collisions which has small nuclear geometric overlap (peripheral), and invariant yields in each case are scaled by a factor to take the different $\langle N_{coll} \rangle$ into account. 
However, in this case the reference production yield of prompt photon will be the one in the central collisions since the magnetic field will be significantly larger in the peripheral collisions.  

The medium mass region, eV to MeV, is excluded by $e^{+}e^{-}$ experiments~\cite{ee_lep}, the high mass region, above GeV, is excluded by the collider experiments~\cite{cms}, and the other regions are considered by some astrophysical arguments~\cite{hb_1, hb_2, bbn_1, bbn_2}. 
Note that the precision of the measured $R_{\rm AA}^{\gamma}$ is the key of this approach and a new phase space in the high mass region, 0.5 to 5 GeV, can be covered. 
Additionally, in some models which the coupling of axion(-like) and gluon is unsuppressed, axion(-like) particles will dominantly decay into hadrons in this mass region~\cite{axion_decay} which means no extra photons will be added in the prompt photon yield.
The expected limits on the coupling constant $g$ in this mass region can go to $10^{-3}$ to $10^{-2}$ GeV$^{-1}$. 
Additionally, this approach can also cover the high mass region of the QCD axion, namely 0.05 to 0.5 GeV, if the contribution of hadronic decay is carefully taken into account, such as the Kim-Shifman-Vainshtein-Zakharov (KSVZ) axion~\cite{ksvz_1, ksvz_2} and Dine-Fischler-Srednicki-Zhitnitsky (DFSZ) axion~\cite{dfsz_1, dfsz_2}. 
This will provide an additional constraint on the coupling of axion(-like) particle to photon. 

\section{Conclusions}
In summary, axion(-like) particles play an important role to solve the most mysterious and interesting puzzles in our universe, namely the missing strong $CP$ violation processes and the origin of dark matter. 
There are many experiments using the properties of photon-axion(-like) particles conversion to search for the axion(-like) signal and push the limits in the extremely low mass region. 

In this paper, we propose an alternative way to search for axion(-like) particles via the prompt photon production in heavy-ion collisions and this approach can provide new constraints in the medium-high mass region where has not been covered before. 
Since an extremely strong magnetic field can be generated in the non-central collisions and photon won't affect by the QGP medium, the nuclear modification factor of the prompt photon production, $R_{\rm AA}^{\gamma}$, can be used to determined the conversion probability of photon to axion(-like) particles. 
Then, this probability can transfer to the upper limit of the coupling constant $g$. 
In other words, the precision of the $R_{\rm AA}^{\gamma}$ measurements of prompt photon production is the key of this search. 
Simple estimations using the future heavy-ion collisions configurations, namely the Au+Au collisions with 100 and 250 TeV center of mass energy, shows that a new phase space, the medium-high mass region, 0.5 to 5 GeV, can be covered.
If the precision of the $R_{\rm AA}^{\gamma}$ can be achieved to 1\% level and the $BL$ factor is 54.5 $T\cdot m$, the upper limit on the coupling constant $g$ can be $10^{-3}$ GeV$^{-1}$. 

\section*{Acknowledgments}
We thank National Cheng Kung University for their support. This work was supported in part by the Ministry of Science and Technology of Taiwan and Higher Education Sprout Project from Ministry of Education of Taiwan.




  

\end{document}